\documentclass[doublecol]{epl2}

\title{A consensus opinion model based on the evolutionary game}

\author{Han-Xin Yang\thanks{E-mail: \email{hxyang01@gmail.com}}} \shortauthor{Han-Xin Yang \etal}

\institute{
   Department of Physics, Fuzhou University, Fuzhou 350116, China}


\abstract{We propose a consensus opinion model based on the
evolutionary game. In our model, both of the two connected agents
receive a benefit if they have the same opinion, otherwise they both
pay a cost. Agents update their opinions by comparing payo?s with
neighbors. The opinion of an agent with higher payoff is more likely
to be imitated. We apply this model in scale-free networks with
tunable degree distribution. Interestingly, we find that there
exists an optimal ratio of cost to benefit, leading to the shortest
consensus time. Qualitative analysis is obtained by examining the
evolution of the opinion clusters. Moreover, we find that the
consensus time decreases as the average degree of the network
increases, but increases with the noise introduced to permit
irrational choices. The dependence of the consensus time on the
network size is found to be a power-law form. For small or larger
ratio of cost to benefit, the consensus time decreases as the degree
exponent increases. However, for moderate ratio of cost to benefit,
the consensus time increases with the degree exponent. Our results
may provide new insights into opinion dynamics driven by the
evolutionary game theory. }

\begin{document}
\maketitle

\section{Introduction}

The dynamics of opinion sharing and competing and the emergence of
consensus have become an active topic of recent research in
statistical and nonlinear physics~\cite{physics}. One of the most
successful methodologies used in opinion dynamics is agent-based
modeling. The idea is to construct the computational devices (known
as agents with some properties) and then simulate them in parallel
to model the real phenomena. In physics this technique can be traced
back to Monte Carlo simulations. A number of agent-based models have
been proposed, which include the voter model~\cite{voter1},
the majority rule-model~\cite{majority1,majority2}, the
bounded-confidence model~\cite{confidence} and the social impact
model~\cite{social1}. Some models display a disorder-order
transition~\cite{order1,order2,order3,order4,order5,order6}, from a
regime in which opinions are arbitrarily diverse to one in which
most individuals hold the same opinion. Other models focus the
emergence of a global
consensus~\cite{consensus1,consensus2,consensus3,consensus4,consensus5,consensus6},
in which all agents finally share the same opinion.

In this Letter, we propose an opinion model based on the
evolutionary game. Evolutionary game theory as a powerful
mathematical framework, has been widely used to understand
cooperative behavior~\cite{cooperation1,cooperation2}, traffic
flow~\cite{traffic1,traffic2}, epidemic
spreading~\cite{epidemic1,epidemic2} and so on. However, to the best
of our knowledge, the impact of evolutionary games on the formation
of public opinion has received little attention.

In fact, it is natural for us to consider different opinions as different
strategies. Individuals with different opinions compete with each
other and gain the corresponding payoffs. We assume that both of the two connected individuals receive a
benefit if they have the same opinion, otherwise they both pay a
cost. The above assumption has been demonstrated by many
psychological experiments in which dissent often leads to punishment
either psychologically or financially, or both, as human individuals
attempt to attain social conformity~\cite{dissent1,dissent2}.

In previous models, an individual usually follows the opinion of a
randomly chosen neighbor~\cite{voter1} or the majority
opinion in the neighborhood~\cite{majority1,majority2}. However, in
our model an individual updates its opinion by comparing payoffs
with a randomly selected neighbors. The more payoff of the chosen
neighbor has, with the higher probability of its opinion will be
imitated. Interestingly, we find that there exists an optimal ratio
of cost to benefit, leading to the shortest consensus time.

\section{Model}

\begin{figure*}
\begin{center}
\scalebox{1}[1]{\includegraphics{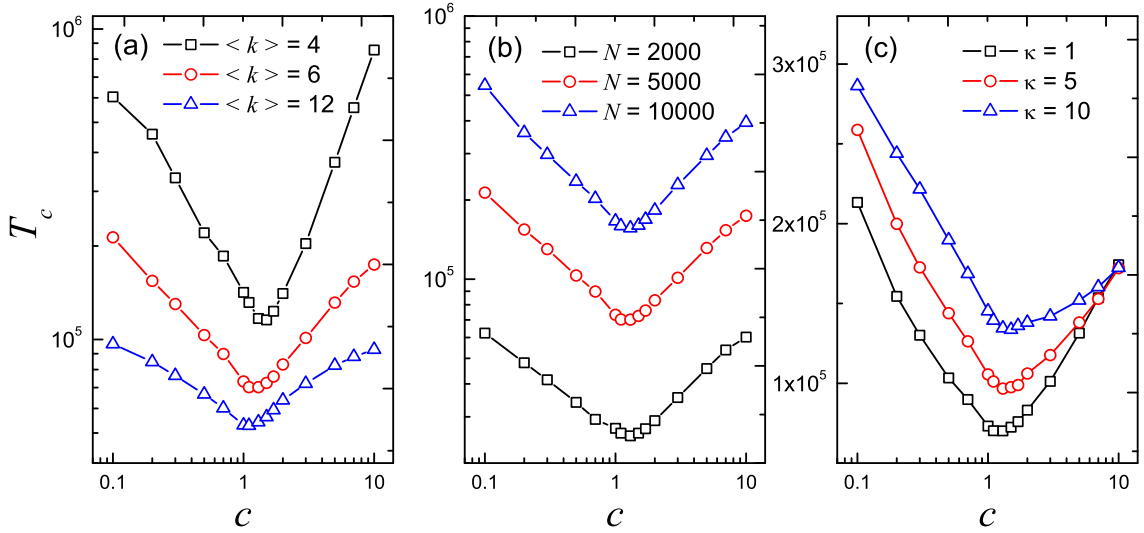}} \caption{(Color
online) (a) The consensus time $T_{c}$ as a function of the cost $c$
for different values of the average degree of the network $\langle k
\rangle$. The network size $N=5000$ and the noise $\kappa=1$. (b)
The consensus time $T_{c}$ as a function of the cost $c$ for
different values of the network size $N$. The average degree of the
network $\langle k \rangle=6$ and the noise $\kappa=1$. (c) The
consensus time $T_{c}$ as a function of the cost $c$ for different
values of the noise $\kappa$. The network size $N=5000$ and the
average degree of the network $\langle k \rangle=6$.} \label{fig1}
\end{center}
\end{figure*}

Our model is described as follows. For a given network of any
topology, each node represents an agent. Initially the two opinions
denoted by the values $\pm 1$ are randomly assigned to agents with
equal probability. Both of the two connected agents receive a
benefit $b$ if they have the same opinion, otherwise they both pay a
cost $c$. Thus, the total payoff of agent $x$ can be calculated as
\begin{equation} \label{1}
P_{x}=\frac{(b-c)k_{x}}{2}+\frac{(b+c)\sum_{i\epsilon\Omega_{x}}S_{x}S_{i}}{2},
\end{equation}
where $k_{x}$ is the degree of agent $x$, $S_{x}$ is the opinion of
agent $x$, and the sum runs over the nearest-neighbor set
$\Omega_{x}$ of agent $x$.

Agents asynchronously update their opinions in a random sequential
order. At each time step, we randomly select an agent $x$ who
obtains the payoff $P_{x}$ according to Eq. (1). Then we choose one
of agent $x$'s nearest neighbors at random, and the chosen agent $y$
also acquires its payoff $P_{y}$ by the same rule. We suppose that
the probability that agent $x$ adopts agent $y$'s opinion is given
by the Fermi function~\cite{feimi1,feimi2,feimi3,feimi4}:
\begin{equation}\label{4}
W(S_{x}\leftarrow S_{y})=\frac{1}{1+\exp[(P_x-P_y)/\kappa]},
\end{equation}
where $\kappa$ characterizes the noise introduced to permit
irrational choices.

In the Fermi updating rule, the
opinion of an agent with higher payoff is more likely to
be imitated. For $P_y>P_x$ ($P_y<P_x$), the probability that agent $x$ adopts agent $y$'s opinion is larger (smaller) than 0.5.

\begin{figure}
\begin{center}
\scalebox{0.6}[0.6]{\includegraphics{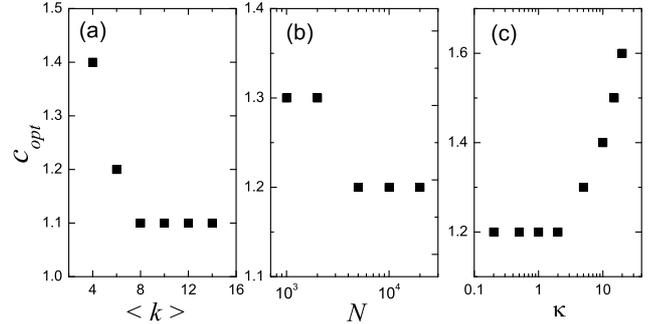}}\caption{(a) The
value of $c_{opt}$ as a function of the average degree of the
network $\langle k \rangle$. The network size $N=5000$ and the noise
$\kappa=1$. (b) The value of $c_{opt}$ as a function of the network
size $N$. The average degree of the network $\langle k \rangle=6$
and the noise $\kappa=1$. (c) The value of $c_{opt}$ as a function
of the noise $\kappa$. The network size $N=5000$ and the average
degree of the network $\langle k \rangle=6$.} \label{fig2}
\end{center}
\end{figure}

\section{Results}

In all following simulations, we use the Barab\'{a}si-Albert
scale-free networks~\cite{BA} to study opinion dynamics with the
evolutionary game rules. Each data point is based on 100
realizations of the network and 10 realizations on each network. To
simplify, we set the benefit from the common opinion as $b=1$.

We define the consensus time $T_{c}$ as the time steps required to
reach the global consensus where all agents in a network share the
same opinion. Figure~\ref{fig1} shows that $T_{c}$ as a function of
the cost $c$ for different values of the average degree of the
network $\langle k \rangle$, the network size $N$ and the noise
$\kappa$. From Fig.~\ref{fig1}, one can see that for given values of
other parameters, there exists an optimal value of $c$, hereafter
denoted by $c_{opt}$, resulting in the shortest consensus time
$T_{c}$.

The value of $c_{opt}$ changes with the average degree of the
network $\langle k \rangle$, the network size $N$ and the noise
$\kappa$. For given values of other parameters, $c_{opt}$ decreases
from 1.4 to 1.1 as $\langle k \rangle$ increases from 4 to 14 [see
Fig.~\ref{fig2}(a)]. The value of $c_{opt}$ decreases from 1.3 to
1.2 as $N$ increases from 1000 to 20000 [see Fig.~\ref{fig2}(b)],
and $c_{opt}$ increases from 1.2 to 1.6 as $\kappa$ increases from
0.2 to 20 [see Fig.~\ref{fig2}(c)]. From Fig.~\ref{fig2}, one can
also observe that the value of $c_{opt}$ is more than the benefit
which is previously set to be 1.

Initially, the two competing opinions are randomly distributed among
the population with the equal probability. Thus, according to the mean-field theory, the initial payoff
of agent $x$ can be approximatively expressed as
$P_{x}(0)=(b-c)k_{x}/2$. For $b>c$, the initial payoff of agent $x$
increases with its degree. However, for $b<c$, $P_{x}(0)$ decreases
as the agent's degree increases.

\begin{figure}
\begin{center}
\scalebox{0.75}[0.75]{\includegraphics{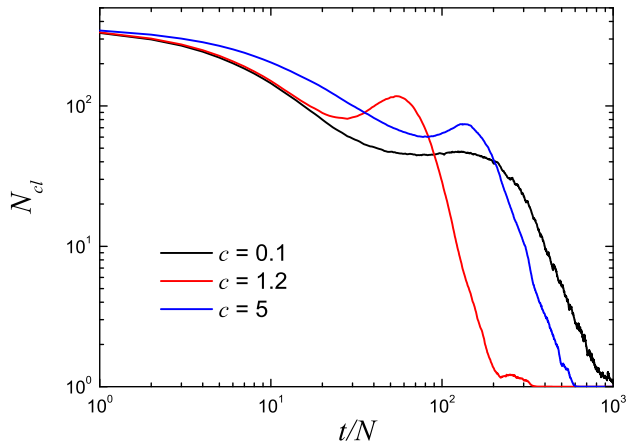}}\caption{(Color
online) The number of opinion clusters $N_{cl}$ as a function of
the rescaled time $t/N$ for different values of the cost $c$. The
network size $N=5000$, the average degree of the network $\langle k
\rangle=6$ and the noise $\kappa=1$.} \label{fig3}
\end{center}
\end{figure}

In scale-free networks a few nodes have high degrees (usually are
called as hubs), while most nodes are of low degrees. It has been
shown that the opinion dynamics is proceeded by formation of some
large opinion clusters centered at
hubs~\cite{cluster1,cluster2,cluster3,cluster4}. A opinion cluster
is a connected component (subgraph) fully occupied by nodes holding
the same opinion. Through the competition of different opinion
clusters, one cluster will invade the others and finally dominate
the system with a global consensus. For very large values of the
cost $c$, initially the payoffs of high-degree agents are much lower than
that of low-degree agents and the opinions of hubs have a very small
probability to be imitated. Thus it becomes difficult for agents to
form large opinion clusters centered at hubs. On the other hand, for very small values
of $c$, a hub has so strong influence on its low-degree neighbors that
the cluster formed by this hub is extremely stable. As a result, the merging of
different clusters become very difficult for too large $c$, leading
to a longer consensus time. Taken together, we can expect that the
shortest consensus time would be realized at the moderate value of
$c$.

To verify the above analysis, we study the number of opinion
clusters $N_{cl}$ as the rescaled time $t/N$ evolves for different
values of the cost $c$. From Fig.~\ref{fig3}, one can see that
initially there exist hundreds of opinion clusters in the network
but $N_{cl}$ eventually decreases to 1 as the time evolves. For
$c=5$, initially $N_{cl}$ decreases much more slowly, compared with
$c=0.1$ and 1.2, indicating that it is hard to form big opinion
clusters when $c$ is large. For $c=0.1$, $N_{cl}$ decreases faster
than the cases of $c=1.2$ and $c=5$ in the early stage ($t/N<100$). However,
when only a few clusters remain in the system, for example,
$N_{cl}<40$, $N_{cl}$ decreases very slowly for $c=0.1$, indicating that the competition
among big opinion clusters becomes furious when $c$ is too small.

\begin{figure*}
\begin{center}
\scalebox{1}[1]{\includegraphics{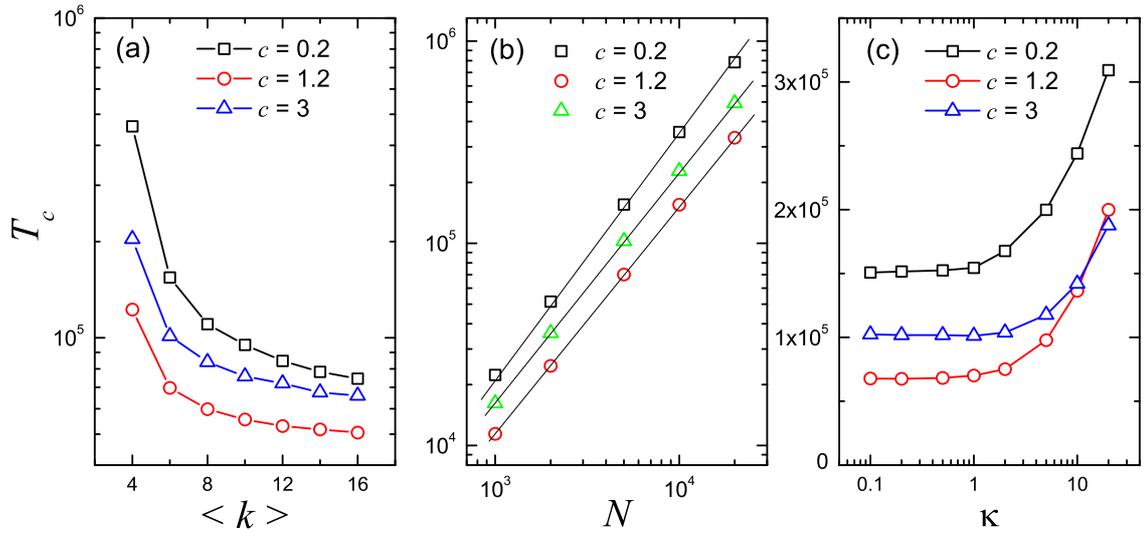}} \caption{(Color
online) (a) The consensus time $T_{c}$ as a function of the average
degree of the network $\langle k \rangle$ for different values of
the cost $c$. The network size $N=5000$ and the noise $\kappa=1$.
(b) The consensus time $T_{c}$ as a function of the network size $N$
for different values of the cost $c$. The average degree of the
network $\langle k \rangle=6$ and the noise $\kappa=1$. The fitted
line has the slope 1.19, 1.13 and 1.15 for $c$ = 0.2, 1.2 and 3
respectively. (c) The consensus time $T_{c}$ as a function of the
noise $\kappa$ for different values of the cost $c$. The network
size $N=5000$ and the average degree of the network $\langle k
\rangle=6$.} \label{fig4}
\end{center}
\end{figure*}

Finally, we investigate the effects of the average degree of the
network $\langle k \rangle$, the network size $N$ and the noise
$\kappa$ on the consensus time. From Fig.~\ref{fig4}(a), we see that
for a given value of the cost $c$, the consensus time $T_{c}$
decreases as $\langle k \rangle$ increases. The consensus time
$T_{c}$ scales as $N^{\beta}$ with the exponent $\beta$ depending on
the value of $c$ [see Fig.~\ref{fig4}(b)]. The exponent $\beta$
=1.19, 1.13, 1.15, correspond to $c$ = 0.2, 1.2, 3 respectively. In
particular, the optimal $c=1.2$ results in the lowest value of
$\beta$. From Fig.~\ref{fig4}(c), we also observe that the consensus
time $T_{c}$ increases as the noise $\kappa$ increases. This
phenomenon indicates that the more rational choice (following the
opinion of the agent with higher payoff) will accelerate the
formation of consensus.

\section{Conclusions and discussions}

In conclusion, we have proposed an opinion model based on the
evolutionary game. An agent receives a benefit if it has the same
opinion with a neighbor, otherwise it pays a cost. An agent randomly
selects a neighbors as a reference. The agent has a higher (lower)
probability to imitate the opinion of the chosen neighbor if its
payoff is lower (larger) than that of the selected neighbor. The results in
scale-free networks show that, the shortest consensus time can be
obtained when the cost of conflicting opinions is a litter larger than
the benefit of common opinions. For very high ratio of cost to
benefit, initially hubs have so low payoffs that their opinions are
seldom imitated, which prevents the formation of large opinion
clusters. On the other hand, too low ratio of cost to benefit makes
the merging of different clusters become very difficult. Thus the shortest consensus time must be realized
at the moderate ratio of cost to
benefit.

The interplay between the opinion dynamics and the evolutionary
games is a very interesting topic. Previous studies have shown that
the introduction of opinion dynamics can greatly affect the
evolution of
cooperation~\cite{interplay1,interplay2,interplay3,interplay4,interplay5}. For
example, Szolnoki and Perc discovered that the spatial selection for
cooperation is enhanced if an appropriate fraction of the population
chooses the most common rather than the most profitable strategy
within the interaction range~\cite{interplay1}. Yang $et.$ found that cooperation is
promoted by punishing neighbors with the opposite strategy~\cite{interplay2}.
Together previous and our works offer an underlying connection
between the evolutionary games and the opinion dynamics.

\acknowledgments This work was supported by the National Natural
Science Foundation of China under Grant No. 6140308 and the training plan for Distinguished
Young Scholars of Fujian Province, China.

\end{document}